\documentclass[aps,prl,preprint,groupedaddress,longbibliography]{revtex4-1}
\usepackage{graphicx,subfigure,amsmath,xcolor}

\begin{document}
\title{Computer simulations of colloidal gels: how hindered particle rotation affects structure and rheology}

\author{Hong T. Nguyen}
\email{hong.nguyen@utdallas.edu}
\affiliation{Department of Materials Science and Engineering, University of Texas at Dallas, Richardson, Texas 75080, United States}

\author{Alan L. Graham}
\affiliation{Department of Mechanical Engineering, University of Colorado -- Denver, Denver, CO USA}

\author{Peter H. Koenig}
\affiliation{Beauty Care Modeling and Simulation, Mason Business Center, 8700 Mason-Montgomery Rd., Mason, OH 45040, USA}

\author{Lev D. Gelb}
\affiliation{Department of Materials Science and Engineering, University of Texas at Dallas, Richardson, Texas 75080, United States}

\date{\today}

\begin{abstract}
  The effects of particle roughness and short-ranged non-central forces on colloidal gels
are studied using computer simulations in which particles experience a sinusoidal
variation in energy as they rotate.
The number of minima $n$ and energy scale $K$ are the key parameters; for large $K$ and $n$,
particle rotation is strongly hindered, but for small $K$ and $n$ particle rotation is nearly free.
A series of systems are simulated and characterized using fractal dimensions,
structure factors, coordination number distributions, bond-angle distributions and linear rheology.
When particles rotate easily, clusters restructure to favor dense packings. This leads to longer gelation times and gels with strand-like morphology. The elastic moduli of such gels scale as $G' \propto \omega^{0.5}$ at high shear frequencies $\omega$. In contrast, hindered particle rotation inhibits restructuring and leads to rapid gelation and diffuse morphology. Such gels are stiffer, with $G'\propto\omega^{0.35}$. The viscous moduli $G''$ in the low-barrier and high-barrier regimes scale according to exponents $0.53$ and $0.5$, respectively. The crossover frequency between elastic and viscous behaviors generally increases with the barrier to rotation. These findings agree qualitatively with some recent experiments on heterogeneously-surface particles and with studies of DLCA-type gels and gels of smooth spheres.

\end{abstract}
\maketitle

\section{Introduction}
A particle gel is a heterogeneous stress-bearing space-spanning network of interacting particles. 
Particle gels can be found in a wide array of practical applications, ranging from tissues engineering \cite{Wang2008} to drug delivery \cite{Xia2005} to biomaterials \cite{Guvendiren2012} to consumer products \cite{Gallegos1999}.
A suspension of colloids can aggregate to form such a network via gelation, a process in which the colloid volume fraction $\phi$ and the inter-particle attraction strength $U$ and range $\Delta$ are among the most important factors. In many cases, the resulting network is self-similar and is characterized by a mass fractal dimension $d_f$. Even though $d_f$ does not fully determine all gel properties, it is still a critical metric and controls the scaling of many properties with volume fraction.

Colloid gelation has been studied computationally and theoretically using a variety of methods. Many models have been developed to explore the microstructure and mechanics of particle aggregates, which can be divided into two families: hard-potential and soft-potential-based models.  In hard-potential models only contact forces are present.  Hard-potential simulations typically make use of Monte Carlo methods, in which particles or aggregates (clusters of bonded particles) are stochastically displaced, avoiding overlapping configurations, and bond irreversibly and rigidly upon collision. Such simulations may be performed either on an discrete underlying lattice\cite{Meakin1983, Gimel1999, Kolb1983, Gimel1995} or continuously (off-lattice) \cite{Fry2002, Hasmy1994, Hasmy1993, Pierce2006, Rottereau2004}. Some off-lattice simulations use Brownian dynamics (BD) instead of Monte Carlo \cite{Bijsterbosch1995, Whittle1997, Whittle1997a, Rzepiela2002, DArjuzon2003} in which realistic diffusive dynamics are included via a Langevin type equation. Studies of hard-potential models have been instrumental in understanding the kinetics of aggregation, fractal properties and cluster structures. Nearly all work to date has focused on spherical particles.

Studies of hard-sphere aggregation models have generally focused on one of two kinetic limits: diffusion-limited cluster aggregation (DLCA) \cite{Weitz1984a}, in which colliding particles always stick together, and reaction-limited cluster aggregation (RLCA) \cite{Weitz1985a} in which particles bond only with a (low) probability upon collision. DLCA is considered a realistic model for colloidal systems with very strong interparticle attractions ($U \gg k_BT$), such as gold \cite{Weitz1984a} or silica nanoparticles \cite{Lin1989}.  RLCA describes systems in which particles must cross an energy barrier before bonding, most often a solvent-induced repulsive force \cite{Jin1996}. By introducing a parameter controlling the probability of bond formation upon collision \cite{Meakin1988} both DLCA and RLCA simulations can be performed with the same computer code. The fractal dimensions of DLCA and RLCA gels are different; $1.78$ for DLCA and $2.1$ for RLCA \cite{Meakin1983}.

Hard-sphere simulations have certain limitations.  First, the bonding in such models is irreversible (bonds never break once formed.)  This fails to describe many technologically important cases with low and intermediate $U$.  Second, the aggregates formed are rigid; the energy of the system is not a continuous function of the volume or simulation cell parameters, and so one cannot directly extract moduli and rheological information from such simulations.
Some MC-type simulations have been augmented with loop-deflection and related ``moves'' {\cite{Jullien1997,ma:etal2001,ma:prevost:scherer2002}} which allow for a limited degree of gel restructuring, but overall such models are unsuitable for studies of long-time aging or mechanical deformation. Finally, hard-sphere simulations can form gels at arbitrarily low volume fractions, which is generally not observed in experimental systems (although real gels with volume fraction below 0.001 have been prepared in some cases.) \cite{Manley2004,Manley2005a,Cipelletti2000}

  More realistic simulation models allow for restructuring, e.g. effect of cluster deformation, bond extension, rotation about bonds, and intra-cluster cluster motion \cite{Meakin1988}. The development of reversible models with breakable bonds allowed for closer contact between simulation and experiment \cite{Kolb1986}. Stochastic bond breakage can be included in MC simulations via a breakage probability \cite{Shih1987}, or bond lifetime\cite{Delgado2004}. In BD simulations, bonds break when their length exceeds a (preset) maximum value \cite{Gelb2007, Whittle1997, Park2015}. Such modifications allowed for investigation of systems with lower $U$ and improved description of the gel network structure and dynamics. They have explained the fractal dimension changes observed in experiment \cite{Liu1990}.

  Models based on soft potentials have continuous forces and can be used to study long-time evolution, rheology and flow behavior in gels. Soft potentials can be either central (acting only on particle centers) or non-central. In studies dealing with central interactions \cite{Zia2014, Griffiths2017, DArjuzon2003}, there is no resistance to angular rotation \cite{Potanin1995a} because the energy only depends on inter-particle separations. Depletion interactions are an example of a central potential \cite{Prasad2003}. Central interactions are appropriate for smooth spherical particles without site-specific bonding. Such interactions typically produce gels with coarse structures, which may or may not be fractal, and which exhibit significant aging and time-dependent rheology \cite{Zia2014,Hsiao2012,Cipelletti2000}.

  Noncentral interactions may arise from particle anisotropy \cite{Mohraz2005} or from close contacts between rough-surfaced or chemically inhomogeneous particles.
   In some computational studies, non-central bonding forces are included through bonds acting between specific points on particle surfaces, which are created dynamically when particles collide. These bonds may be freely orienting \cite{Whittle1997a, Park2015, Dickinson1994} or govered by angular and torsional terms \cite{Gelb2007}. In such simulations the surface points at which the bond acts remain fixed, unless the bond is broken and a new bond is formed. Models of this type are more computationally complex than central-force models, both because of the data-management associated with the dynamic creation and removal of bonds, and because the bond forces in the simulation now act on particle surfaces, introducing torques and complicating the calculation of stresses and other quantities. This class of model has been successfully used to study gels in which the bonds are due to specific chemical interactions \cite{Gelb2007} or due to the interaction of surface-bonded polymers \cite{Whittle1997a, Dickinson1994}. This approach is less suitable for rough-surfaced particles, in which it may be possible for one particle to ``roll'' around on the surface of another without losing contact.
 
  Another approach to incorporating non-central interactions is the use of ``patchy'' models \cite{Kern2003}, in which each particle's surface is decorated with interaction sites at fixed positions\cite{Zaccarelli2007}. Wang \textit{et al.} \cite{Wang2019} recently studied such a model, where each particle had $42$ randomly arranged interaction sites. This surface heterogeneity shifted the gel point away from that predicted by Baxter's isotropic model \cite{Baxter1968}, and increased the elastic modulus of the colloidal suspension. Because of their complex structure, only small systems of $500$ particles were simulated.
  In an alternative approach, Del Gado~{\emph{et al.}} introduced non-central forces by adding  a three-body bending term with a preferred angle\cite{Bouzid2018, Colombo2013, Colombo2014}.  However, it is difficult to physically justify the use of a preferred three-body angle in terms of microscopic interparticle interactions.  The results obtained are only reasonable descriptions of gels of low coordination number, and it is not clear how to extend this approach to systems with larger contact numbers as is commonly seen in experiments. 

  Recently, Pantina and Furst experimentally measured tangential forces between particles in isolated colloidal aggregates, demonstrating the relevance of contact interactions in the rheology and dynamics of DLCA gels \cite{Pantina2005,Furst2007}. Laxton and Berg likewise used bending of linear aggregates to probe the rigidity of interparticle bonds {\cite{Laxton2007}}.
  Many experimental systems composed of rough-surfaced particles \cite{Hsiao2017, Schroyen2019} have been synthesized and characterized; such work has been reviewed recently by Hsiao and Pradeep \cite{Hsiao2019}. Those results emphasize the important role of particle surface chemistry and anisotropy in aggregation kinetics, restructuring and rheology.

  In this paper, we study a simple model for non-central surface-type interactions that can be applied to particles varying from very smooth to very rough.
  Barriers to the rotational motion of bonded particles are accounted for by introducing a sinusoidally varying two-body angular potential $U_a$ with two parameters: the barrier height (energy scale) $K$ and the number of energy minima $n$ explored as one particle completes a rotation. We show that for appropriate choices of $K$ and $n$ the model produces stable gel structures at low volume fractions. The gels obtained are fractal, with $d_f \sim 1.99-2.16$. The dynamics, network structure and linear rheology of systems at various $K$, $n$, and $\phi$ are characterized and compared. The gel time is found to be decreasing with increasing $K$ and  $n$.
  For low barrier height $K$, particles are nearly free to rotate and the resulting colloidal networks resemble the coarse, strand-like structures obtained in simulations of soft central-force models. 
  The scaling exponent of the elastic moduli $G'$ and the frequency is $\sim 0.5$, as in colloidal suspensions of smooth spheres without hydrodynamic interactions. For higher barrier heights and large numbers of minima, particle rotation is restricted, leading to much less intracluster restructuring and more diffuse networks. The rheology for higher barrier heights is also different; the low-frequency limit of the modulus is well-defined and increases with increasing $n$, and at high frequency the scaling exponent is reduced to $0.35$. Transitions from solidlike to liquidlike frequency response are shifted to higher frequencies and are also $n$-dependent. The power-law exponent describing the viscous frequency-dependent moduli $G''$ varies only weakly with $K$.

\section{Model and Methods}
The model gel consists of monodisperse spherical particles which interact through both a non-bonding pair-wise potential and interactions between bonded particles. The parameters chosen and approximations made are appropriate to the case of short, stiff bonds which do not break over the course of the simulation.  
The non-bonding interaction is a truncated-and-shifted Lennard-Jones (LJ) potential:

\begin{equation}
U_{LJ}(r) = 
   \begin{cases}
   4 \epsilon	\left[\left(\frac{\sigma'}{r}\right)^{12} - \left(\frac{\sigma'}{r}\right)^6 + \frac{1}{4}\right] & ~~~~~r \leq r_c \\
   0, & ~~~~~r > r_c \\
   \end{cases}
\label{eq:ULJ}
\end{equation}
$\epsilon$ controls the energy scale, and $r$ is the interparticle distance.  The nominal particle diameter is $\sigma$; the length parameter in $U_{LJ}$ is chosen to be $\sigma' = 2^{-1/6}r_c$, which ensures that only the repulsive core is retained and that the non-bonding interaction term goes exactly to zero at $r_c$. In contrast to core potentials such as the repulsive spherical \cite{Whittle1997} and r-shifted LJ \cite{Gelb2007}, this choice offers both energy and force continuity at $r_c$, which is important for computational stability.

Inter-particle bonds are created dynamically over the course of the simulation.  A new bond is made when two particles which are not bonded to each other approach to a separation shorter than the equilibrium bond length $l_0 = r_c$ . These simulations are therefore ``DLCA-like'' in that there is no energetic barrier to bond formation. The stretching of interparticle bonds is modeled with a harmonic potential:
\begin{equation}
U_b(l) = \frac{1}{2}\epsilon_{b}\left(l -l_0\right)^2
\label{eq:Ub}
\end{equation}
where $\epsilon_{b}$ set the scale of bond energy. The range of short-range attraction in colloidal systems is typically a few percent of the particle size \cite{Poon1997}, here we take $l_0=r_c=1.02\sigma$, which also sets the parameter $\sigma' = 0.90782\sigma$.

Bond breakage, which is important in nonlinear processes such as large-amplitude shear \cite{Park2015} and gel collapse \cite{Buscall2009}, is not considered here; once formed, bonds remain active for the duration of the simulation. The use of an unbreakable harmonic potential will obviously lead to unphysical results under large strains or shears, but is acceptable for small-amplitude rheological tests.

\begin{figure}[htbp]
\includegraphics[width=2.5 in]{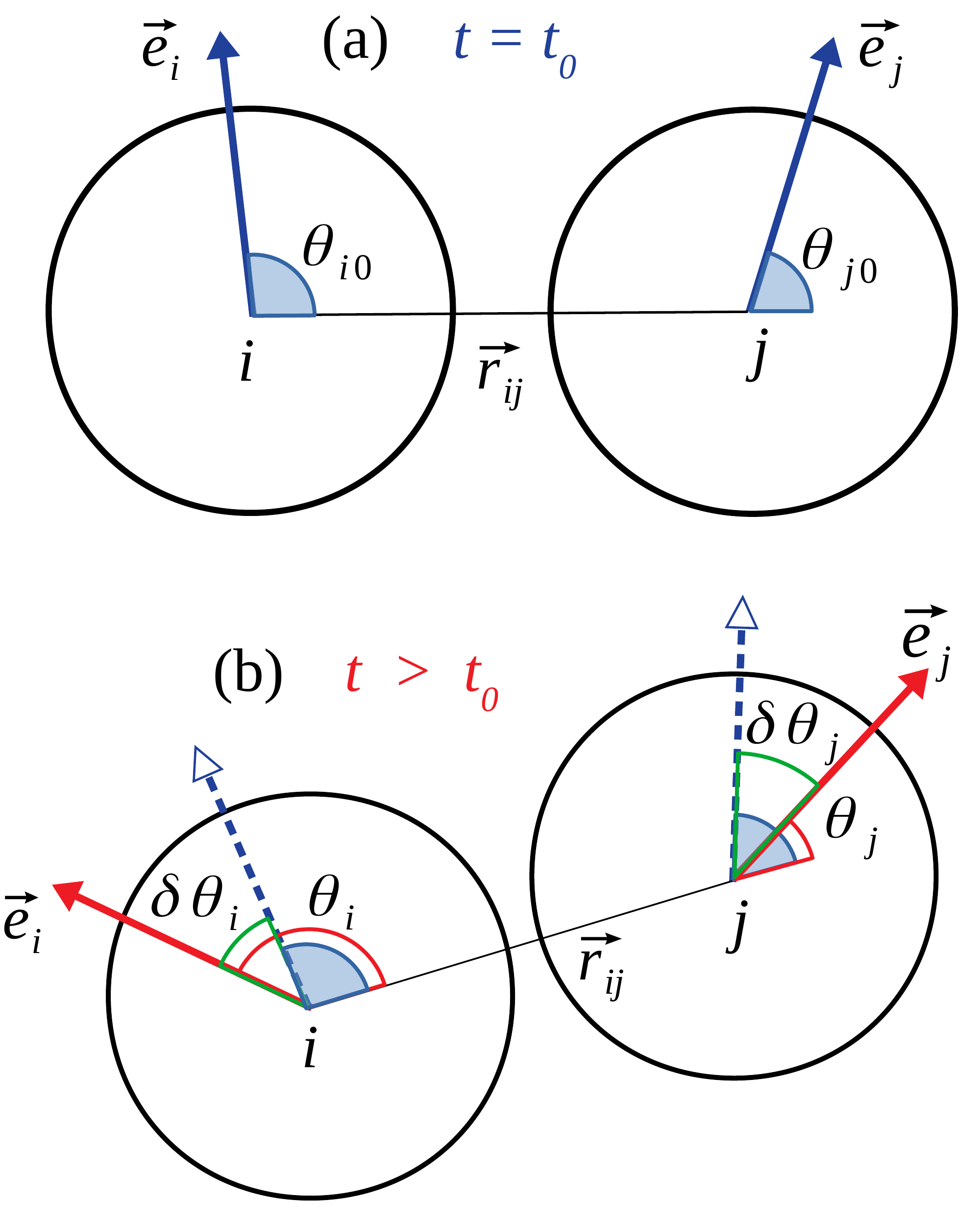}
\caption{Schematic illustration of angular interactions parameters. $\vec{e_i}$ and $\vec{e_j}$ are the internal orientational vectors of $i, j$ respectively. (a) A new bond is created between particle $i$ and $j$ at time $t = t_{0}$. (b) At later time $t > t_{0}$ , $\vec{e_i}$ and $\vec{e_j}$ vary from their initial values.}
\label{fig:model}
\end{figure}

In real gels particles may experience both frictional forces and site-specific interactions when they come into contact. These may be due to inhomogeneity of the particle surfaces; particles may not be perfectly smooth or spherical, and their surfaces may also be chemically inhomogeneous (e.g., hydrogen-bonding sites may be non-uniformly distributed.)  Such effects are likely to be more significant for small (nanoscale) particles.  
In such systems, the angular (tangential) motion of one particle relative to its near (bonded) neighbors will cause a change in potential energy, even though the interparticle distance does not change.  We model this variation with a sinusoidal term $U_{rot}(\delta \theta) = K\left[1-\cos(n\delta \theta)\right]$ (see Fig.~\ref{fig:model}), where $\delta\theta$ is the angular displacement. Here $K$ and $n$ control the frequency and magnitude of the oscillation, such that as a particle completes a rotation the energy goes through $n$ minima separated by barriers of height $2K$. A similar oscillatory term is applied to torsional motion around the interparticle bond.

The specifics of the implementation are as follows. Each particle $i$ is assigned an internal orientation $\vec{e_i}$ fixed in the particle frame of reference. The orientational configuration of a pair of bonded particles $i,j$ is then a triad $ \Gamma_{ij}(t)= \{\theta_i(t),\theta_j(t),\gamma_{ij}(t)\}$, where 
$\theta_i= \cos^{-1}\left[ \Vec{e}_i \cdot \Vec{r}_{ij}/(|\Vec{e}_i| \cdot |\Vec{r}_{ij}|) \right]$ is the angle between the internal vector of particle $i$ and the interparticle direction $\Vec{r}_{ij}$; similar expressions are applied for particle $j$. Torsional orientation is defined by the angle between the two orientational vectors $\Vec{e}_{i}$ and $\Vec{e}_{j}$, $\gamma_{ij} = \cos^{-1} \left[ \Vec{e}_i \cdot \Vec{e}_j/(|\Vec{e}_i| \cdot |\Vec{e}_{j}|)\right]$. Suppose that a bond is formed between particles $i$ and $j$ at time $t_{0}$. As time progresses, the triad will deviate from its initial value $\{\theta_{i0},\theta_{j0},\gamma_{ij0}\}$.
The bond energy $U_a$ between the particles is then a sum of three oscillatory terms as described above, one each for the angular displacements and one for the torsion, with all $K$ and $n$ parameters taken to be the same for simplicity:   
\begin{equation}
U_a(\theta_i,\theta_j,\gamma_{ij}) = 
  K \left[ 3 - \cos \left( n \delta \theta_i \right) - 
  \cos \left( n \delta\theta_j \right) - 
  \cos \left( n \delta\gamma_{ij} \right)\right].
\label{eq:Ua}
\end{equation}
where $\delta\theta_{i}=\theta_{i} - \theta_{i0}$ and  $\delta\gamma_{ij}= \gamma_{ij} - \gamma_{ij0}$.  
Because this potential takes its minimum value of zero at $\delta\theta_{i}=\delta\theta_{j}=\delta\gamma_{ij} = 0$, there are no discontinuous changes in the energy, forces or torques when a new bond is created.

In principle $K$ should vary with the interparticle distance $l$; as particles are moved apart the effect of surface inhomogeneities on their interactions should decrease \cite{Colombo2013,Gelb2007}. However, since for the stiff bonds considered here $l$ will not vary very far from $l_0$, it is reasonable to simply take $K$ independent of interparticle distance.

\subsection{Simulation protocol}
The initial state of the system contains $N$ particles randomly placed in a cubic box, avoiding any overlap. Periodic conditions are applied in all directions. We choose to ignore hydrodynamics, as is common in simulations of this type; we note in this regard that De~Graaf~\emph{et al.} simulated a simple colloid model with only central forces and found that ignoring hydrodynamics did affect gelation dynamics but did not change the structures produced {\cite{DeGraaf2019a}}. The particle motion is thus described by Langevin dynamics \cite{Bijsterbosch1995, Schneider1978}

\begin{equation}
    m\frac{d\Vec{v}_i}{dt}=-\xi_T \Vec{v}_i(t) + \Vec{F}_i(t) + \Vec{F}^R_i(t).
    \label{eq:force}
\end{equation}
$\xi_T = 3\pi \eta \sigma$ is the coefficient that controls the drag force due to the solvent, where $\eta$ is the solvent viscosity.
$\Vec{v}_i$ is the velocity of particle $i$, and $\Vec{F}_i$ is the total pair-wise force on particle $i$. $\Vec{F^R_i}(t)$ is a random force satisfying condition $\langle F^R_{i,\alpha}(t) F^R_{j,\beta}(t') \rangle = 2 \xi_T k_B T \delta_{ij} \delta_{\alpha\beta}\delta(t-t')$, where $\alpha, \beta = x,y,z $ are Cartesian components, $T$ is temperature, and $k_B$ is Boltzmann's constant. $\Vec{F}^R_i$ simulates fluctuating forces exerted on gel particles by the solvent. In the absence of hydrodynamic interactions there is no coupling between rotational and translational motion \cite{Bijsterbosch1995}. Rigid-body rotational motion is described by \cite{Whittle1997}
\begin{equation}
    I\frac{d\omega_i}{dt}=-\xi_R \omega_i(t) + \Vec{Q}_i(t) + \Vec{Q}^R_i(t).
    \label{eq:torque}
\end{equation}
Here $\Vec{Q}_i, \Vec{Q}^R_i$ are the total and random torques on particle $i$, $\xi_R = \pi\eta\sigma^3$ is the rotational friction coefficient and  $I$ is the particle's moment of inertia, which is a scalar for spherical particles.
The random torque satisfies $\langle Q^R_{i}(t) Q^R_{j}(t') \rangle = 2 \xi_R k_B T \delta_{ij}\delta(t-t')$.  The rotational and translational diffusion coefficients $D_R$ and $D_T$ are related by the identity $\sigma^2 D_R/3=D_T=k_B T /3\pi \eta \sigma$. 

Key simulation parameters are summarized in Table \ref{tab:simparams}. Because the particles are monodisperse, they all have the same mass $m$. All quantities are reported in reduced units, with temperature scaled by $\epsilon$, distance scaled by the particle diameter $\sigma$.  The reduced unit for time is $\tau = \sqrt{m\sigma^2 /\epsilon}$, but a more useful choice is the Brownian relaxation time $\tau_R = \sigma^2/4D_T$, which is the time taken for a particle to move a distance equal to its own diameter \cite{Whittle1997}; most temporal quantities are therefore reported in units of $\tau_R$.  $\xi_T$ is set to $10.0 ~m \tau^{-1}$ (thus, $\xi_R = 10/3 ~ m \tau^{-1} \sigma^2$) in all simulations, which is in the range commonly seen in the gel simulation literature\cite{Bouzid2018,DeGraaf2019a}. The integration time step is $25\times 10^{-4}\tau$ in simulations of gelation or low-frequency shear, and $1 \times 10^{-4}\tau$ in simulations of high-frequency shear, discussed further below. 
The real times corresponding to these simulated times depends on the properties of the particles simulated. For near-buoyant 100~nm particles in water at ambient conditions, for instance, $\tau_R$ corresponds to $5.06 \times 10^{-4}$ seconds.
  The simulation cell edge length is $L = 60\sigma$, which is substantially larger than the characteristic length scale of the simulated gels at all volume fractions considered. This ensures that the cell periodicity does not influence the gel structures or rheological results presented below.

\begin{table}[htbp]
  \caption{ Summary of simulation parameters}
\begin{tabular}{llr}
\hline
System size &  $L $  & $60\sigma$ \\
LJ length parameter &  $\sigma' $  & $0.90872\sigma$ \\
System temperature & $T$ & $0.2\epsilon$ \\
Equilibrium bond length & $l_0$ & $1.02\sigma$ \\
Volume fraction & $\phi$ & $0.02-0.075$\\
Time step & $\delta t$ & $25 \times 10^{-4}$ or $10^{-4}\tau$ \\
Translational friction coefficient & $\xi_T$ & 10 $m \tau^{-1}$ \\
Bond energy & $\epsilon_{b}$  & $400\epsilon$ \\
Duration of gelation simulations &  & $2 \times 10^2 - 2 \times 10^4\tau_R$\\
\hline
\end{tabular}
\label{tab:simparams}
\end{table}

All simulation are performed using a modified version of LAMMPS \cite{Plimpton1995} that implements the $U_a$ potential described above. Implementation of $U_a$ within LAMMPS is not straightforward because the triad $\Gamma_{ij}^0 = \{ \theta_{i0}, \theta_{j0}, \gamma_{ij0}\}$ is only determined at the moment that a bond between two particles is created. 
Keeping track of $\Gamma_{ij}$ for each bond must therefore be carried out ``on the fly'', which is accomplished with new data fields added within LAMMPS's base classes. Furthermore, interprocess communication routines were modified so as to keep all processors updated with the $\Gamma_{ij}$ values during the simulation.

\subsection{Characterization of simulated gel structures}

\textit{Gelation time}

To quantify the aggregation process, we monitor the size distribution of clusters formed during the run; a particle is part of a cluster if at least one bond connects it to the cluster. The gelation time $t_{gel}$ is defined as the shortest time at which the largest cluster simultaneously spans all three dimensions of the simulation box.  At $t_{gel}$ there may still be smaller clusters not yet attached to the percolating network. Aggregation is complete at time $t_c$ when only a single cluster remains, though further evolution through internal restructuring may still occur after this time.

\textit{Fractal dimension}

The fractal dimension $d_f$ is a critical measure governing structural and mechanical properties of gels \cite{Weitz1985a,Laxton2007}.
Generally $d_f$ is greater for gels composed of more compact clusters.
On sufficiently large length scales gels are homogeneous, but at intermediate scales they are inhomogeneous with a characteristic length $\xi$, often thought of as an average cluster size. $\xi$ is likewise proportional to the position of a broad minimum at large $r$ in the radial distribution function $g(r)$ \cite{Hasmy1994}, and scales with volume fraction $\phi$ according to 
\begin{equation}
\xi \sim \phi^{-1/(3-d_f)}.
\label{eq:xi}
\end{equation}
Thus one can estimate $d_f$ by locating the minimum in $g(r)$ and plotting it against $\phi$ on a log-log scale.

The structure factor $S(q)$ is also computed from $g(r)$ through
\begin{equation}
S(q) = 1 + \frac{4\pi \rho }{q} \int r[g(r)-1] \sin(qr) dr.
\end{equation}
In fractal gels $S(q) \sim q^{-d_f}$ in the intermediate range of wave vector $q$ corresponding to the real-space intermediate scale defined above.  One can therefore also determine $d_f$ from the slope of $\log[S(q)] \text{ vs. } \log[q]$ in the appropriate range of $q$.

\textit{Rheology}

To determine the rheological characteristics of the simulated gels, we use non-equilibrium simulations with Lees-Edwards boundary conditions \cite{Lees1972}. Computation of the stress components is straightforward since particles interact with differentiable pairwise potentials:
\begin{equation}
    \sigma_{\alpha\beta} = \rho k_B T - V^{-1}\sum^{N}_{i,j=1} r_{\alpha ij}F_{\beta ij} 
\end{equation}
where $V$ is the system volume, $\alpha, \beta = x,y,z$ denotes the Cartesian directions, $\Vec{r}_{ij}$ and $\Vec{F}_{ij}$ are the center-to-center vector and total pair-wise force between particles $i$ and $j$ as defined above.  In this method the gel is subjected to a sinusoidal shear of amplitude $\gamma_0$  and frequency $\omega$ in the $x\-y$ plane:
$
    \gamma (t) = \gamma_0 \sin(\omega t).
$
Because the $x$-dimension remains fixed, this deformation protocol preserves the system volume.
After running the simulation for $m$ cycles of shear, the storage $G'(\omega)$ and loss $G''(\omega)$ are calculated from
\begin{equation}
    G'(\omega) = \frac{\omega}{\pi m \gamma_0} \int^{2m\pi/\omega}_{0} \sigma(t) \sin(\omega t) dt
\end{equation}
\begin{equation}
    G''(\omega) = \frac{\omega}{\pi m \gamma_0} \int^{2m\pi/\omega}_{0} \sigma(t) \cos(\omega t) dt
\end{equation}
As is typically done, the shear frequency is reported as a dimensionless quantity $\alpha = \omega \tau_R$.
We consider only the linear response regime. The shear amplitude is kept small in order to preserve the overall gel structure; also, as in other studies, new bonds are not allowed to form during the shear simulations\cite{Whittle1997}.
To reduce statistical noise, many cycles are simulated.  For $\alpha < 10$ (low frequencies), $10$ cycles and a time step of $25\times 10^{-4}\tau$ were found to give good results. For $10 < \alpha$, $50$ cycles were sufficient, and a time step of $1\times 10^{-4}\tau$ was used in order to accomodate the high strain rate.

\section{Results}
Simulations were performed for models with a range of bond potential parameters $K$ and $n$ at several volume fractions.  In each case two types of simulation are performed: gelation and oscillatory shearing.  The gelation simulations were all run up to $2 \times 10^4 \tau_R$,
which is well past the point of complete aggregation in each case.  In the following discussion we first analyze a particular system in detail and then explore how changes in the bonding potential affect gel properties.

\subsection{Gelation dynamics and gel structure at $K=1$ and $n=1$}

\begin{figure*}[htbp]
\includegraphics[width=\textwidth]{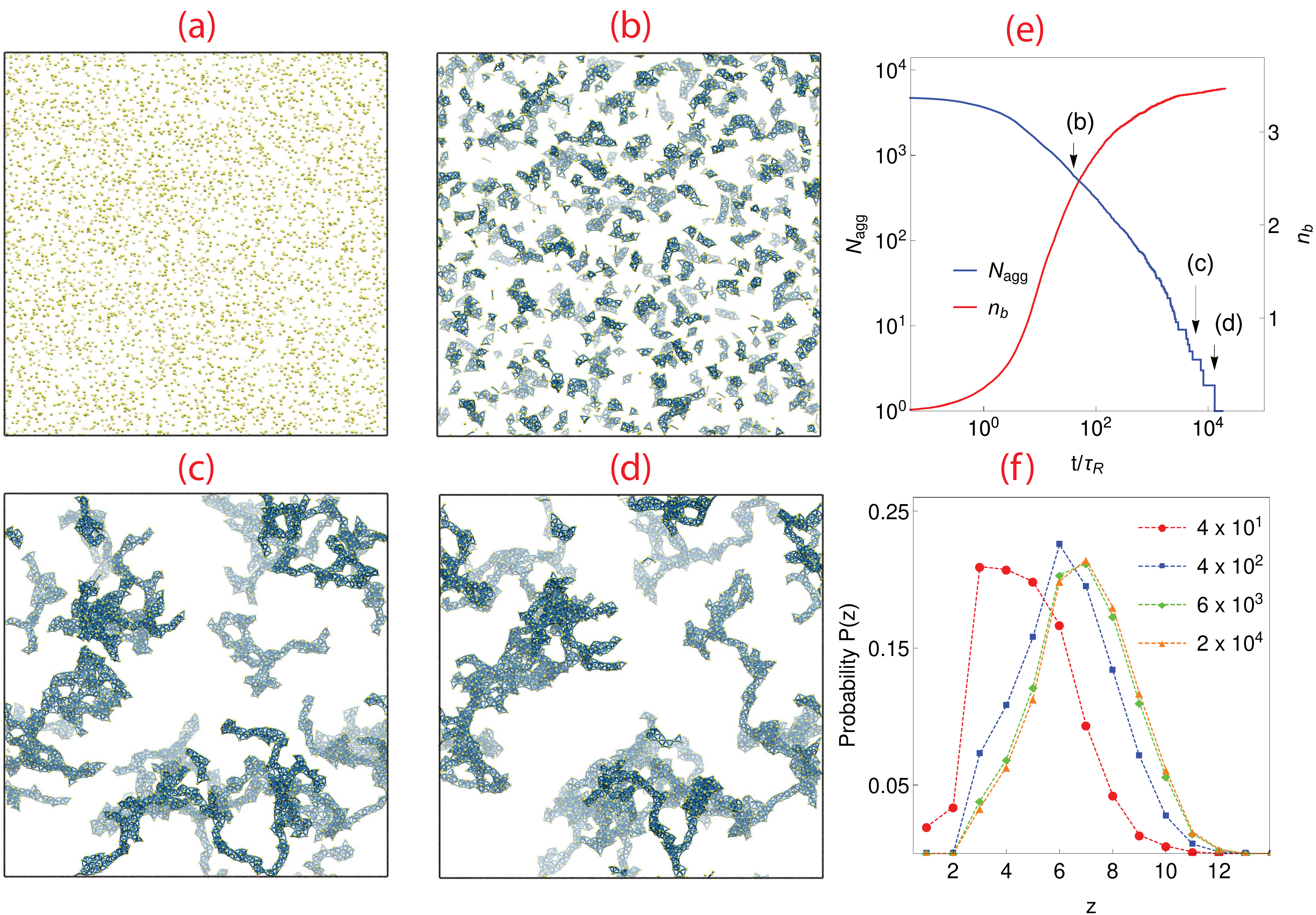}
\caption{Gelation simulation with $\phi = 0.02, K = 0.1$ and $n = 1$. (a) initial state with particles placed at random, (b) at $t = 4\times 10^1\tau_R$, particles have aggregated to form small clusters, (c) at $t_{gel}=6 \times 10^3 \tau_R$, a system-spanning network is formed, (d) complete aggregation at $t_c=1.3 \times 10^4 \tau_R$.  Particles are as yellow dots while bonds are shown as dark blue rods; note that because only a single periodic cell is shown the gel network looks unconnected.  
  (e) time evolution of average bond number $n_b$ and number of aggregate $N_{agg}$, with times corresponding to configurations (b-d) indicated by arrows.
  (f) distribution of contact number $z$ at different times before and after complete aggregation (times given in legend). 
  }
\label{fig:gelation}
\end{figure*}

We begin with an analysis of gelation dynamics in a system with volume fraction $\phi = 0.02$,  $K = 0.1$ and $n = 1$, in which the barrier to particle rotation is very small.  A series of snapshots from the gelation simulation and several quantities describing the bond network are shown in Fig.~{\ref{fig:gelation}}.  
In the initial state particles are randomly scattered, Fig.~\ref{fig:gelation}(a). As they collide, they bond and form clusters, Fig.~\ref{fig:gelation}(b). As clusters grow the dynamics slows because the cluster diffusion constant is inversely proportional to its diameter. The clusters also slowly become more compact. This occurs because the particles can move about within each cluster (as long as no bonds are broken) and form new bonds with other particles in the cluster; the cluster interior becomes denser and locally trapped in this way.

When clusters meet they merge into larger clusters as bonds are formed between particles near their surfaces. At $t_{gel} \sim 6 \times 10^3\tau_R$, in Fig.~\ref{fig:gelation}(c), the largest cluster spans the simulation box, although there are many smaller clusters still present. Aggregation continues with the attachment of smaller structures to the spanning network while cluster compactification is still taking place. Full aggregation is reached at $t_c \sim 1.3 \times 10^4 \tau_R$, Fig.~\ref{fig:gelation}(d).  

Fig.~\ref{fig:gelation}(e) shows the time evolution of the number of clusters present ($N_{agg}$) and the number of bonds per particle, $n_b$.  At early times ($t < 6\tau_R$) the kinetics of aggregation is governed by the particle-particle collision rate, and $n_b$ increases according to a power law.  At $t \sim 6 \tau_R$ the supply of monomers is largely exhausted and cluster-cluster aggregation becomes the dominant growth mechanism. In this region $N_{agg}$ decays according a power law. $n_b$ is growing but no longer according to a power-law; it reaches a plateau at around $t \sim 8 \times 10^2 \tau_R$ after which it only increases a very small amount for the remainder of the simulation.  This behavior suggests that all possible intra-cluster relaxations have occurred by this time; further bond formation only occurs between clusters. The system-spanning network does not form until $t_{gel} = 6\times 10^3\tau_R$. There are still ${\cal{O}}(10)$ clusters present at $t_{gel}$; the $N_{agg}$ plot exhibits large steps after this time as the remaining clusters very slowly collide and merge.
The final structure contains an average of $n_b \sim 3.5$ bonds per atom, indicating a much denser packing than in DLCA networks which have $2.0$ bonds per particle \cite{Hasmy1993,Hasmy1994}.
Note that because bonds cannot be broken in these simulations, the degree to which the gel structure can evolve over long times is limited; in particular, spinodal-decomposition type coarsening is not possible because such large-scale restructuring necessarily involves breaking bonds.

Fig.~\ref{fig:gelation}(f) shows $P(z)$, the distribution of the number of bonds $z$ made by a particle, at various times. $z$ strongly affects the rigidity and mechanical response of the gel \cite{Hsiao2012, Zia2014}. In simulations of colloidal gel aging Zia et al.\cite{Zia2014} showed that particles on the surface of network strands adopt $1 \leq z \leq 3$, while particles in the interior have  $z \geq 8$. As time progresses, the peak $z_{max}$ of $P(z)$ clearly shifts to higher $z$, suggesting network coarsening by additional bond formation. At $t= 4 \times 10^1\tau_R$, $P(z)$ has a maximum at $z=3$, implying most particles are near cluster surfaces. At $t= 4 \times 10^2 \tau_R$ the maximum in $P(z)$ shifts to the isostatic value $z_{iso}=6$ \cite{Hsiao2012}, consistent with the thickening of network strands visible in Fig.~\ref{fig:gelation}(c) and (d). This trend persists until the gel point, at which the maximum of $P(z)$ occurs at $z=7$, consistent with the final value of $n_b \sim 3.5$ found in Fig.~\ref{fig:gelation}(e). There is no further significant change in $P(z)$ over the remainder of the simulation. 

\subsection{Effects of $n$ and $K$ on dynamics and gel structure}
\begin{figure*}[htbp]
\includegraphics[width=\textwidth]{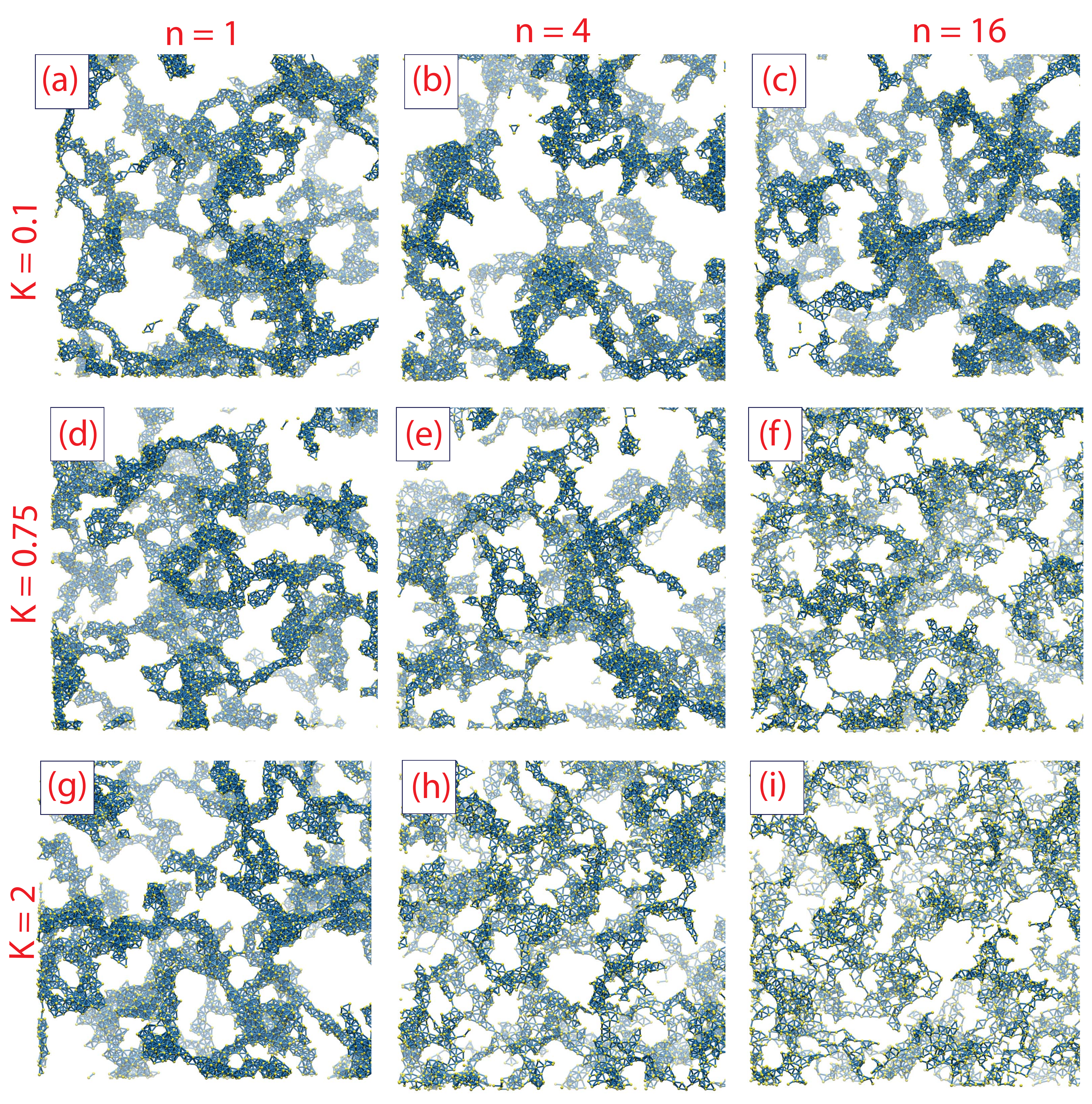}
\caption{Effect of $K$ and $n$ on final gel morphology for systems $\phi = 0.045$, snapshots taken after complete aggregation. Top row (a,b,c) $K = 0.1$, middle row (d, e ,f) $K = 0.75$ and bottom row (g, h, i) $K = 2$, from left to right $n = 1,4,\text{ and } 16$ respectively.}
\label{fig:arraysnap}
\end{figure*}
Both local structure and overall gel morphology can be tuned by adjusting the parameters $K$ and $n$.  Fig.~\ref{fig:arraysnap} shows final gel structures from simulations with low $K = 0.1$ ($K/k_BT = 0.5$), moderately high $K = 0.75$ ($K/k_BT=3.75$) and very high $K = 2$ ($K/k_BT=10$), for each of $n = 1, 4$ and $16$; each structure shown is that observed at the time of complete aggregation, rather than at the end of the gelation simulation.

With $K = 0.1$, varying $n$ has very little effect on gel structure. This is as expected; when the barrier height $K$ is smaller than thermal activation $k_B T$, particles easily move between minima, and so changing the number of minima is unlikely to have a significant effect. In other words, when the angular potential $U_a$ is small compared with the bonding potential $U_b$, we recover behavior typical of models with strong short-range attraction but without angular rotation, viz. references \cite{Zia2014} and \cite{Griffiths2017}.

With $K = 0.75$ there is a visible change in the texture of the gel with increasing $n$. At low $n$ the gel strands are thick (coarse) and very similar to those in the $K=0.1$ gels, but at high $n$ the gel has thinner strands and a finer texture. At small $n$ the local minima in the orientational potential are much broader than at high $n$, so that even though escape from a minimum is kinetically limited there is still the possibility of significant orientational motion and intra-cluster restructuring. As a result, small clusters become compact before aggregation and the gel texture resembles that obtained at low $K$.  For higher $n$ the minima in the orientational energy are narrow and prevent intra-cluster restructuring, so clusters do not become compact before they aggregate.

At $K=2$ the possibility of thermal escape from an orientational potential minimum is extremely small. The effect of $n$ on the gel structure is similar to that observed at $K=0.75$ but more pronounced; at small $n$ some local restructuring can still occur, resulting in relatively thick strands, but at $n=4$ and especially at $n=16$ much finer textures are obtained. In particular, at $n=16$ there is almost no compaction into strands and the structure closely resembles the DLCA models produced in stochastic simulations of aggregating hard spheres \cite{Gelb2019}.

\begin{figure*}[htbp]
\includegraphics[width=5.5 in]{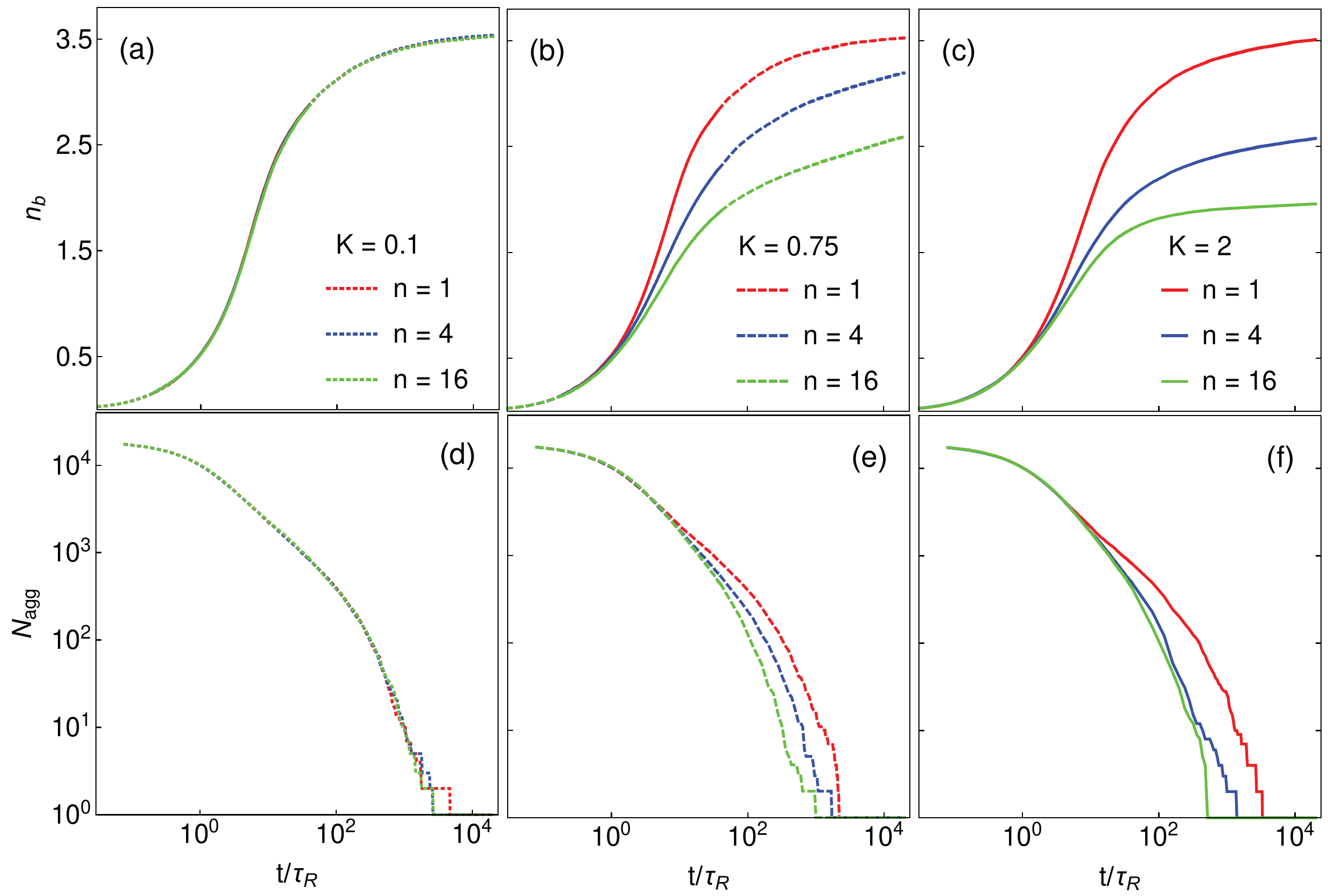}
\caption{Time evolution of the average bond number $n_b$ (a, b, c) and the number of aggregate $N_{agg}$ (d, e, f) during gelation for various systems $L = 60\sigma, \phi = 0.045$. For each $K$, results for three $n$ are presented: $n = 1$ (red), $n = 4$ (blue), $n = 16$ (green).}
\label{fig:dynamics}
\end{figure*}

The kinetics of gelation and compaction are likewise influenced by $K$ and $n$.  Figs.~\ref{fig:dynamics}(a--c) show the time evolution of $n_b$ for the systems in Fig.~\ref{fig:arraysnap}. For $K = 0.1$, $n_b(t)$ is essentially independent of $n$, while at higher $K$ $n_b$ this is not the case.  The bond
creation rate at short times is almost independent of $K$ and $n$, as it is controlled primarily by the diffusion rate of monomers and very small clusters (dimers, trimers, etc.) for which internal restructuring is largely irrelevant. For $K=0.75$, $n_b$ reaches a plateau for the $n=1$ system, but slow restructuring is still occuring at the longest times simulated $n=4$ and $n=16$. In this system the probability of thermal escape from orientational local minima is small but not extremely so, which means that restructuring can still occur over long time scales.  For $K=2.0$ such restructuring is much slower. In this case, only the $n=1$ and $n=4$ systems are still creating small numbers of new bonds at the end of the simulation. At $K=2.0$ and $n=16$ the $n_b(t)$ curve is quite flat and nearly equal to 2.0 at late times, implying DLCA-type structure \cite{Hasmy1993,Hasmy1994}.

\begin{figure}[htbp]
\includegraphics[width=2.0 in]{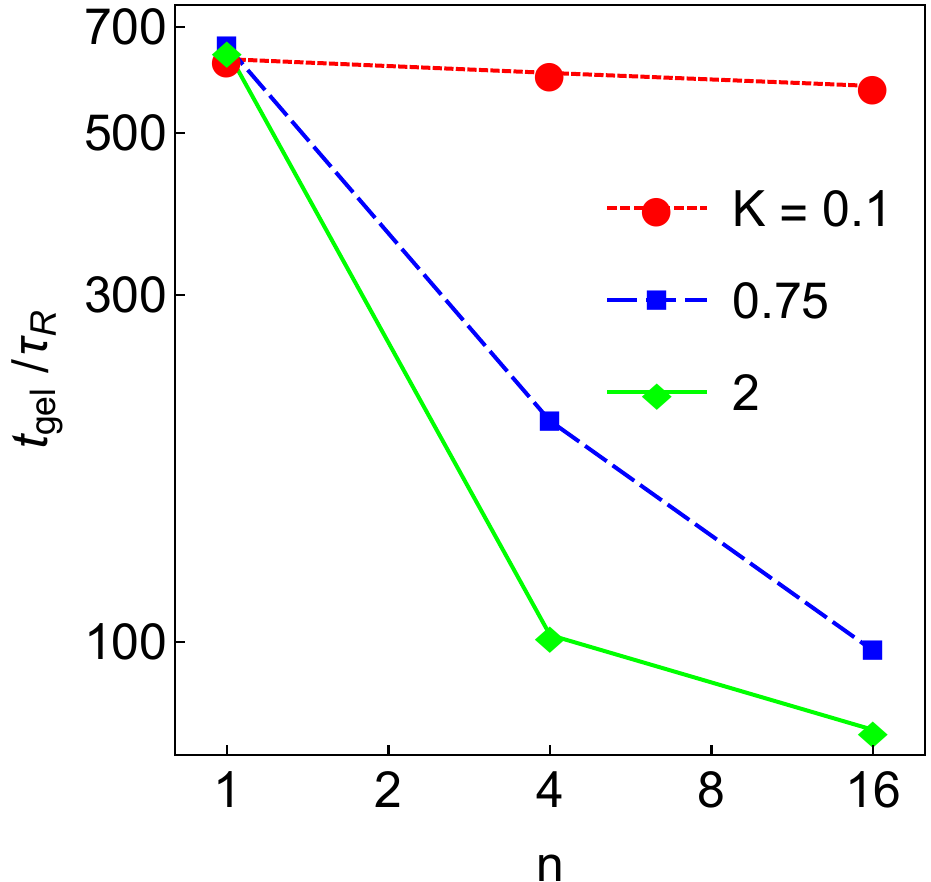}
\caption{Gel time $t_{gel}$ as a function of $n$ for the simulated systems in Fig.~\ref{fig:arraysnap}, plotted on a log-log scale.  Each data point is an average taken over $8$ independent simulations, with a relative standard error of less than $10\%$. Lines are included only as guides to the eye.}
\label{fig:tgel}
\end{figure}

The aggregation dynamics as quantified by $N_{agg}$ are shown in Fig.~\ref{fig:dynamics}(d--f). 
$N_{agg}$ decays slowly at short times, crosses a `shoulder' at intermediate time and then approaches $1$ at late times. $N_{agg}(t)$ is independent of $K$ and $n$ at short time ($t < 2\tau_R$) where particle-particle aggregation dominates, similar to $n_b(t)$.  As in the case of $n_b$, with $K=1$ the $N_{agg}(t)$ is independent of $n$ at all times.  At higher $K$, $N_{agg}$ decreases faster for higher $n$.  This occurs because in these systems clusters and aggregates are more diffuse (highly branched) at higher $n$ (see Fig.~\ref{fig:arraysnap}), which makes them larger and more likely to come into contact with each other. 
As a result, the aggregation kinetics are faster in these systems.  (Equivalently, if clusters can restructure to become more compact, they will be smaller and less likely to collide, which slows down the aggregation process.)  These effects result in a substantial dependence of the gel time $t_{gel}$ on $n$, which is shown in Fig.~\ref{fig:tgel}; the gel time decreases with $n$ in all cases, but most dramatically at high $K$.  At intermediate $K=0.75$, further increase in $n$ may still affect gelation kinetics, while at $K=2$ further increases in $n$ seem unlikely to substantially reduce the gel time.  More generally, these data suggest that any substantial degree of surface roughness or rotational friction will substantially decrease the gel time.

\begin{figure*}[htbp]
\includegraphics[width=5.5in]{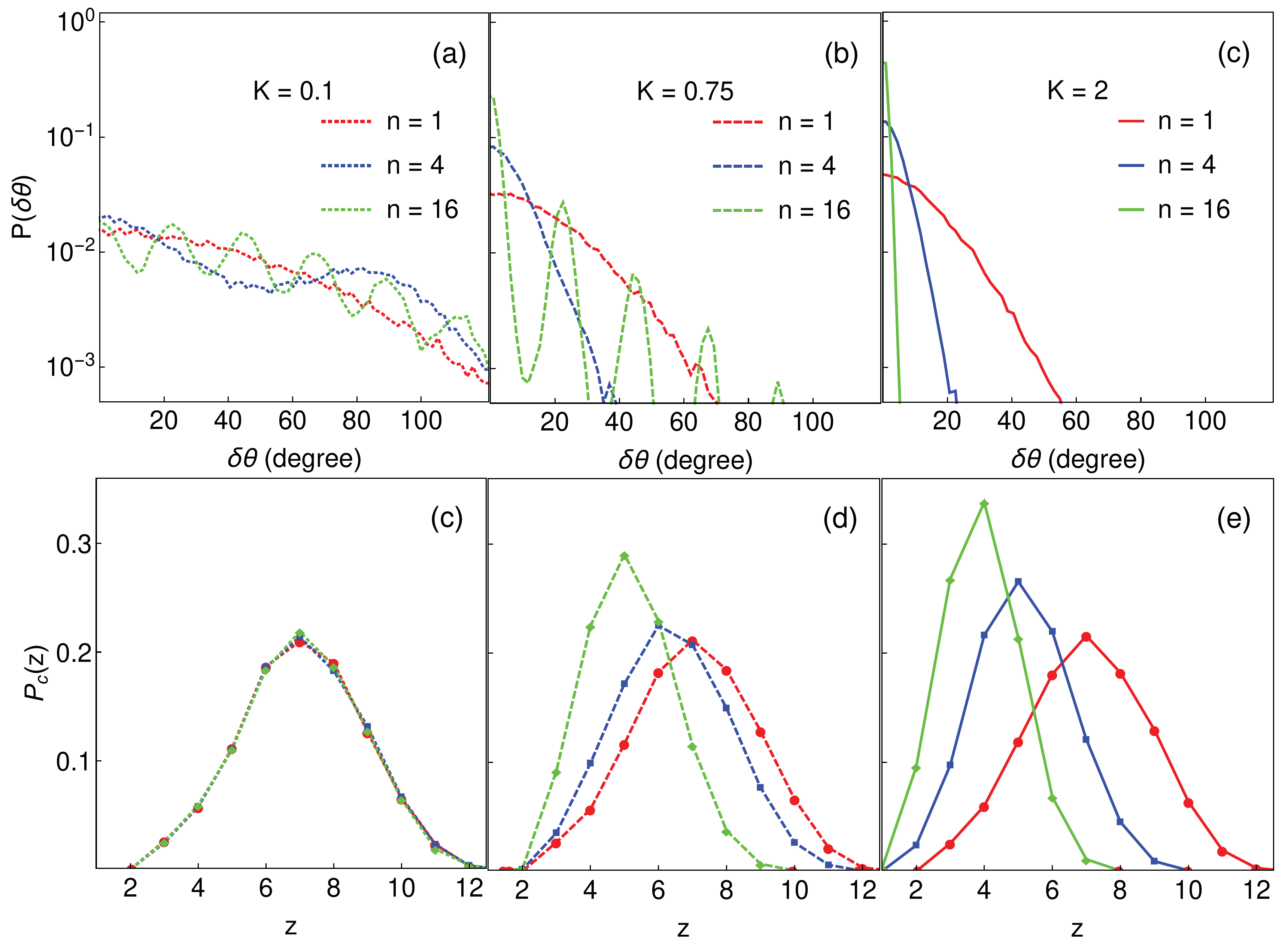}
\caption{Bond angle and contact number distributions in the final gel states from Fig.~\ref{fig:arraysnap}. Colors are the same as in Fig.~\ref{fig:dynamics}. Top row: distributions of angular deviation; only $\delta\theta \geq 0$ is shown because of symmetry. Bottom row: contact number distributions.} 
\label{fig:zdist}
\end{figure*}

We now discuss qualitative analyses of the structures of the final gel states displayed in Fig.~\ref{fig:dynamics} and how they depend on $K$ and $n$.
Fig.~\ref{fig:zdist}(a--c) show the angular deviation distributions $P(\delta\theta;K,n)$, and Fig.~\ref{fig:zdist}(d--f) show the contact number distributions. The angular term $U_a$ has $n$ minima located at $ \pm  m 360^ \circ / n$ (with $m=0,1,2...n/2$), independent of $K$. Depending on $K$, some of these minima are well-populated in the gel structure, resulting in peaks in $P(\delta\theta;K,n)$.  In particular, many such peaks are visible for $K = 0.1 \text{ and } 0.75$, but only the $m=0$ peak is visible for the $K=2$ systems.  Overall, the $m = 0$ maximum is the most intense in all cases, because this is the minimum corresponding to the initial orientation at which each bond is formed; peaks at larger $m$ are populated only through orientational motion of the bonded particles.

At $K = 0.1$, all the minima are well-populated for each of $n =1, 4 \text{ and } 16$, consistent with a large degree of restructuring in these systems. In each case, the intensity of peaks at larger $m$ decays logarithmically with $m$. Even for $n=1$, where there is only a single minimum, the angular distribution is quite broad, indicating that particles have substantially re-oriented within that minimum.  For $K = 0.75$, at $n=1$ and $n=4$ only a single peak is visible (and narrower than for the $K=0.1$ cases), while at $n=16$ only a few peaks at small $m$ have high intensity, corresponding to population only of minima near to the orientation at bond formation.  In other words, large deviations from initial contact angle ($\delta\theta  \gtrsim 100^{\circ}$) are not observed at $K=0.75$.  At $K=2.0$, only the $m=0$ minimum is populated for each $n$; the particles are clearly unable to rotate into neighboring minima.

These results can be understood in term of diffusion. After a new bond is created, $\delta\theta$ starts to deviate from its initial value of zero due to thermal motion and the forces exerted by other particles. For each bond, a particle can explore its $m=0$ minimum or it can hop over the potential barrier into neighboring minima, and from there into other minima.  The hopping rate proportional to the barrier height and width, i.e. $\propto n^{-1} \exp(-K/k_BT) $. Thus, as $K$ rises the probability to cross the barrier decreases, resulting in a narrower achievable range of $\delta\theta$. It is clear from these data that the significant restructuring and densification noted earlier for low $K$ and $n$ is made possible by a large degree of rotational motion of particles within clusters; conversely, large $K$ and $n$ values prevent rotational motion within clusters and lead to more diffuse local structure.

The effect of $n$ and $K$ on cluster compactification are also reflected in the contact number distribution $P_c(z)$. As $n$ and $K$ are increased the maxima in $P_c(z)$ are shifted to lower values, consistent with lower-coordinated, less-compact structures. For $K = 0.1$, $P_z(c)$ is independent of $n$ and symmetrically distributed about $z=7$. A significant portion of particles have $z > 8$, implying thick and close-packed network strands \cite{Zia2014}.  At higher $K$, the effect of increasing $n$ is to reduce the number of high-$z$ particles and increase the number of low-$z$ particles, making $P_c(z)$ asymmetric. This effect is stronger for higher $K$. For example, with $K=2$ and $n=16$, more than $70\%$ of particles have $\leq 4$ contacts, consistent with highly branched and diffuse fractal structures as observed in Fig.~\ref{fig:arraysnap}(f). 

\begin{figure*}[htbp]
\includegraphics[width=5.5in]{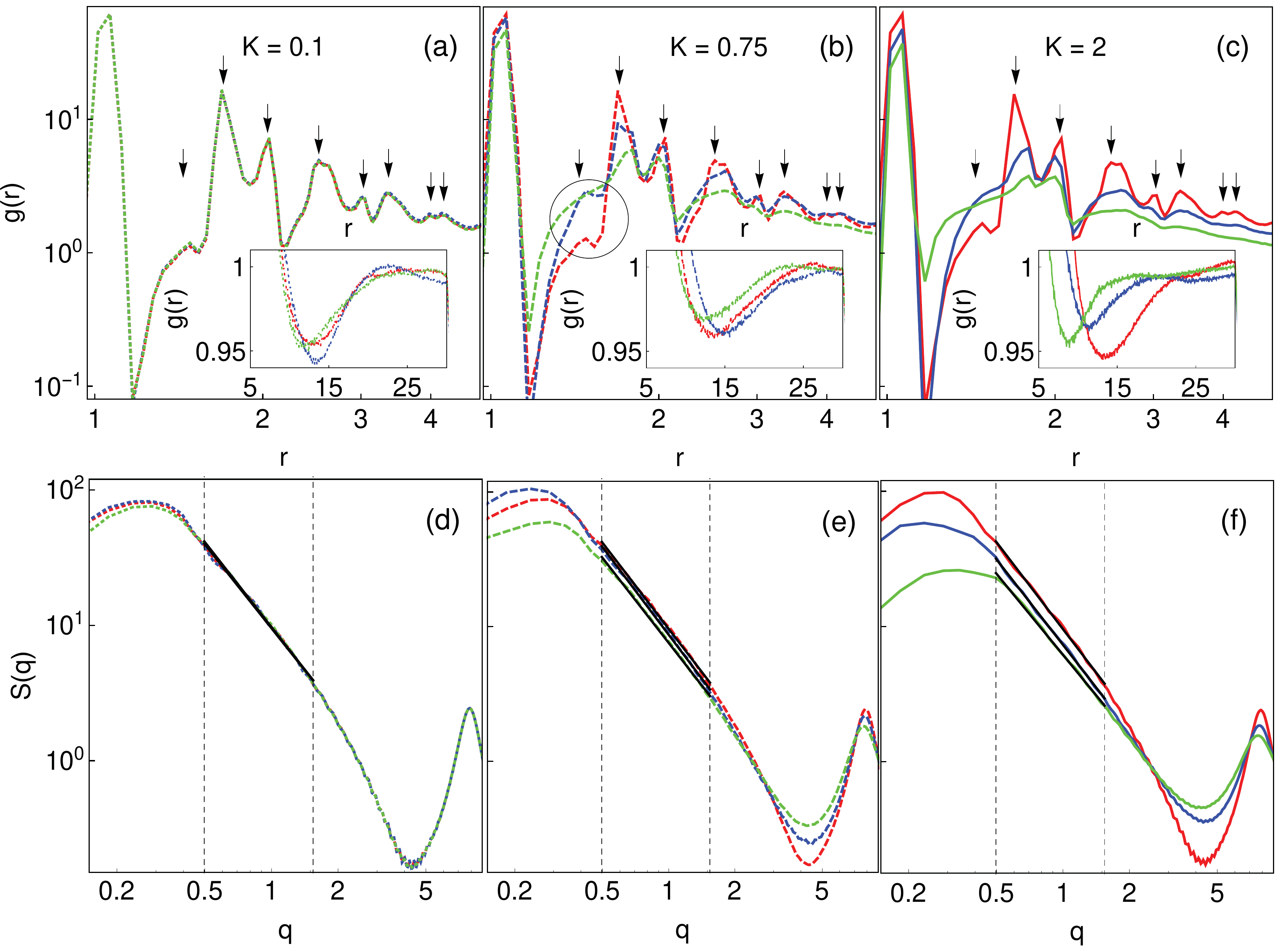}
\caption{Spatial distribution functions in the final gel states from Fig.~\ref{fig:arraysnap}. Colors are the same as in Fig~.\ref{fig:dynamics}. Top row: radical distribution functions, $g(r)$, with peak positions indicated by solid arrows.  Bottom row: static structure factors $S(q)$. Insets in (a--c) highlight the minima in $g(r)$ at large $r$. The solid lines are fits to the fractal region as used to obtain $d_f$. The dashed vertical lines indicate the fitting region. $g(r)$ and $S(q)$ curves are averages over $8$ independent simulations. The circle in panel (b) highlights the $r = 1.4$ peak.}
\label{fig:rdfsofq}
\end{figure*}

Fig.~\ref{fig:rdfsofq}(a--c) shows the radial distribution function $g(r)$ for the final gel states for $K$ and $n$ as in Fig.~\ref{fig:zdist}. All samples have a sharp peak at $r \sim 1.0$ corresponding to the first-nearest neighbor contact. For $K = 0.1$ ( panel (a)), there is only a very weak dependence on $n$. Peaks are observed at $r = 1.41, 1.73, 2.5, 3, 3.35, 4$ and $4.2$; these positions are present in face-centered close-packed structures. This behavior is consistent with the compactified appearance of all $K=0.1$ samples observed earlier.  At $K=0.75$ and $K=2.0$, the $g(r)$ curves vary with $n$. At these $K$ values, the intensity of almost all structure in $g(r)$ is diminished with increasing $n$.  The exception to this behavior is at $r=1.41$, where the intensity of $g(r)$ increases with increasing $n$. This $r$ corresponds to the second-nearest neighbor distance in a square-planar configuration of particles, which is a more open structure than the tetrahedral packing signified by the $r=1.73$ peak.  This behavior is thus consistent with the development of more diffuse gels with increasing barriers to local restructuring.

The corresponding static structure factors are shown in Fig.~\ref{fig:rdfsofq}(d--f).  All curves have a strong peak at low wavevector $q$. For $K=0.1$, the data for different $n$ are essentially identical, while at higher $K$ there is some $n$-dependence. As $n$ is increased the intensities of the low-$q$ peak and the oscillation at high $q$ are reduced, suggesting a more homogeneous mass distribution corresponding to a more diffuse gel structure.

\begin{table}[htbp]
  \caption{Fractal dimensions $d_f$ of the final states of systems with $\phi = 0.045$. $d_f$ was measured by a linear fit to the plot $\log[S(q)] \text{ vs } \log(q)$ restricted to the fractal region $\Delta q \sigma^{-1} =[0.50, 1.54]$, which is the linear portion of data shown in Fig.~\ref{fig:rdfsofq}(d, e, f). 8 independent simulations were performed for each $K$ and $n$. $d_f$ is extracted separately for each realization, then the results are averaged. The standard error given is the standard deviation of the best fit $d_f$ values obtained from linear regression, divided by the square root of the sample size (8).
}
  \begin{tabular}{|l|c|c|c|}
\hline
$K$ &$d_{f, n = 1}$  &   $d_{f, n = 4}$   &  $d_{f, n = 16}$  \\
\hline
0.1    &  2.05 $\pm$ 0.02   &   2.05 $\pm$ 0.01   &  2.08  $\pm$ 0.02 \\
0.75  &  2.12 $\pm$ 0.01   &   2.16 $\pm$ 0.02   &  2.10  $\pm$ 0.02 \\
2.0    &  2.13 $\pm$ 0.03   &   2.10 $\pm$ 0.02   &  1.99  $\pm$ 0.02 \\
\hline
\end{tabular}
\label{tab:df}
\end{table}

The dependence of $d_f$ on $K$ and $n$ is reported in Table \ref{tab:df}, where $d_f$ is extracted from the slope of the linear portion of the plot $S(q) \text{ vs. } q$ on log-log scale. In general, $d_f$  depends only weakly on $n$ for $K = 0.1$ and $0.75$.  At high $K$, $d_f$ decreases slightly with increasing $n$, consistent with the finer structure observed in many other ways. That such a trend in $d_f$ is only observed at high $K$ even though there are clearly changes in gel structure with $n$ at lower $K$ supports the notion that $d_f$ is not an unique measure that fully characterizes the gel structure. For the largest $K$ and $n$ considered $d_f = 1.99$, which, while lower than the fractal dimensions obtained under all other conditions, is still significantly higher than the commonly quoted value of $d_f = 1.78$ for DLCA gels  \cite{Weitz1985a,Lin1989}. This suggests that even under these conditions local restructuring still has some effect on gel structure. We have repeated these simulations and analysis at $\phi = 0.025$ and obtained fractal dimensions in quantitative agreement with this given in  Table \ref{tab:df} (data not shown).

\subsection{Effects of $K$ and $n$ on rheology}

\begin{figure*}[htbp]
  \includegraphics[width=\textwidth]{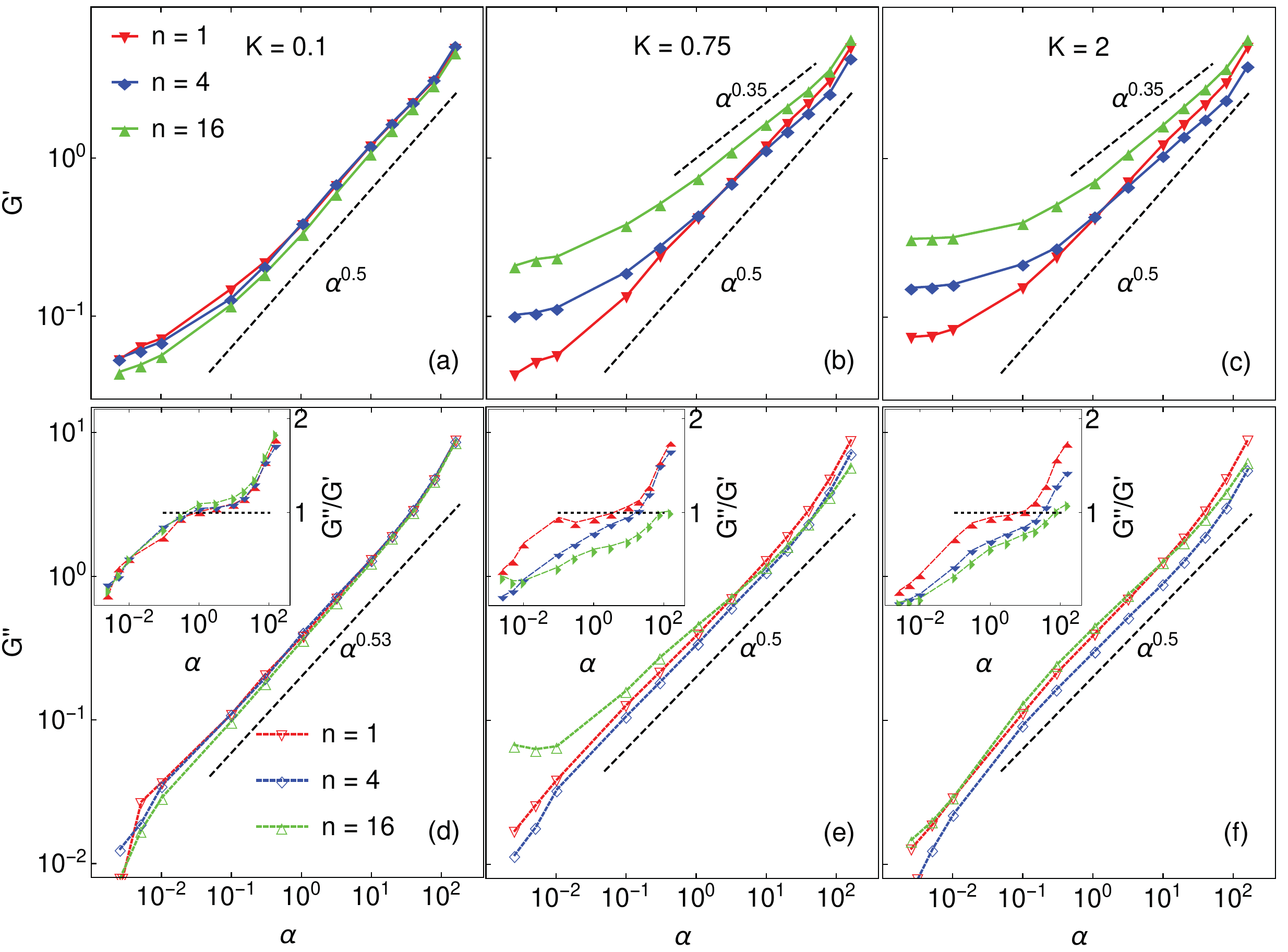}
  \caption{ Storage moduli $G'$ (solid symbols) and loss moduli $G''$(open symbols) for systems with $\phi = 0.075$ with various $K$ and $n$, plotted against dimensionless frequency.  (a, d) $K = 0.1$, (b, e) $K = 0.75$ and (c, f) $K = 2$. The dashed lines indicate different power laws. Insets in (d-f) give the ratio $G''/G'$. The horizontal dotted lines in the inset mark the transition from solid-like to liquid-like behaviors.}
\label{fig:g}
\end{figure*}

Next we turn to the rheological behavior of the model gels. The data presented is taken at volume fraction $\phi = 0.075$, which is higher than the $\phi=0.045$ systems discussed in the preceding section.  Selected simulations performed at $\phi=0.045$ did show the same general trends and scaling behavior as those at $\phi=0.075$, but the results were quite noisy.  Increasing the volume fraction to $\phi=0.075$ raises both the storage and loss moduli by approximately one order of magnitude, which greatly improves the statistical quality of these data. For 100~nm near-buoyant colloidal particles, the unitless frequency range from 0.005 to 100 corresponds to a real frequency range of $\omega = 9.88$ to $1.97\times 10^5$ s$^{-1}$.

As noted earlier, results are reported in terms of the dimensionless frequency $\alpha = \omega \tau_R$.
The elastic modulus $G'(\alpha)$ and the loss modulus $G''(\alpha)$ are measured over the frequency range $ 2 \times 10^{-2} < \alpha < 2 \times 10^2$.
The results are shown in Fig.~\ref{fig:g} and all show the following features commonly observed in
experiments on DLCA-like colloidal gels \citep{Trappe2000a,Prasad2003}.  At low $\alpha$ ($< 10^{-1}$), $G'$ approaches a frequency-independent limiting value, $G'_{0}$, which here depends on both $n$ and $K$. The loss modulus $G''$ is likewise smaller than the elastic modulus $G'$; all these systems behave as elastic solids.  As $\alpha$ increases, $G''$ increases faster than $G'$. Finally, both $G'$ and $G''$ exhibit a power-law scaling with exponents that depend on both $K$ and $n$ .

For $K = 0.1$, varying $n$ has no effect on $G'$ except at the lowest frequencies studied, where there is a small change in the limiting value.  At high $\alpha$ $G'$ scales as $\alpha^{0.5}$ in all cases, in accord with the findings of Zia~{\emph{et al.}} in simulations of gels of attractive hard-sphere colloids \cite{Zia2014}, albeit at rather higher volume fraction.  We note that the same scaling $\alpha^{0.5}$ has elsewhere been reported for short-ranged attractive colloidal suspensions in the ``free draining limit'' of negligible hydrodynamic interactions. \cite{Lionberger1994,Swan2014,Varga2015,Elliott1998}.  Because the same exponent is observed in the gel networks here, it suggests that the network backbone of $K = 0.1$ systems plays a secondary role in transmitting the stress, i.e. the stress correlation length scale is short-ranged.

There are two limiting regimes of rheological behavior, separated by the frequency $\alpha_c$ such that $G'(\alpha_c) = G''(\alpha_c)$. At low $\alpha$ gels are elastic-dominated or solid-like, and at high $\alpha$ they are viscous-dominated or liquid-like.  At $K=0.1$ these systems have nearly identical $\alpha_c \sim  1$, independent of $n$ (see inset in Fig.~\ref{fig:g}(d)), such that the crossover occurs when the rate of perturbation is comparable with the Brownian relaxation time. $G''$ varies with $\alpha$ as $\alpha^{0.53}$, except at very low and very high frequencies where there are small deviations.

Markedly different rheological behaviors arise at higher $K$.
At $K = 0.75$ the high-$\alpha$ elastic responses for $n=4$ and $n=16$ deviate from the $\alpha^{0.5}$ law; their scaling exponents are $0.4$ and $0.35$, respectively. These are significantly weaker than the $n = 1$ system which still follows the $\alpha^{0.5}$ power law. This weaker dependence on frequency signals the network structure becoming more rigid and more particles moving in response to the external perturbation.
  The crossover frequencies are well separated and increase with $n$, with $\alpha_c \sim 3$, $15$ and $76$ for $n = 1$, $4$, and $16$, respectively.
At $K = 2$, the scaling exponent $0.35$  is seen for both $n = 4$ and $n=16$.  As for $K=0.75$, the crossover frequency increases with increasing $n$, $\alpha_c \sim 6, 28$ and $66$ for $n = 1, 4$ and $16$ respectively. A well-defined low-frequency plateau of $G'$ is clearly observed for all systems at $K=2$, while for lower $K$ this is not these case. The low-frequency limit of $G'$ for $K=2.0$ varies with $n$; these systems become stiffer with increasing $n$. 

  The power-law scaling of $G''$ is only weakly dependent on $n$ and $K$. It varies from $0.53$ for $K = 1$ to $0.5$ for higher $K$, independent of $n$.  In all but one case, only small deviations from power-law scaling are observed over the entire range of frequency studied. However, for $K=0.75$ and $n = 16$, $G''$ appears to be almost frequency-independent at low $\alpha$. This appears to be due to the gel aging during the shear simulations \cite{Cipelletti2000, DArjuzon2003}, even though the creation of new bonds is suppressed \cite{Whittle1997}.  At this $K$ and $n$, the barriers to rotational motion are not very high and are narrow enough that some particles jump to neighboring minima and do not return to the original state at the end of each shear cycle.  This is confirmed by calculating $G''$ values from individual successive shear cycles, which increase systematically. In all the other systems studied, there was no significant variation of rheological properties over the duration of the shearing simulations.

The results in Fig.~\ref{fig:g} demonstrate that gels stiffen as the resistance to inter-particle rotation increases, even though the gel texture becomes finer and there are fewer total bonds formed.  As $n$ and $K$ increase, the network connectivity increasingly becomes more important in the elastic response, while the stress correlation length grows.
It is interesting to note that the same stiffening has recently been observed in experiments on colloid suspensions\cite{Schroyen2019}. High-frequency rheology was used to investigate the effect of heterogeneous particle surfaces controlled by varying the thickness of the stabilizing layer on particle surfaces. That study found scaling with $\alpha^{0.5}$ for smooth particles in the intermediate frequency range $10^0  <\alpha < 10^3$. However, for higher surface heterogenities the elastic responses showed a weaker power-law dependence, in agreement with our findings.

The data in Fig.~\ref{fig:g}(a) suggests that for $K=0.1$ and $K=0.75$ a 
a much lower probing frequency is required in order to obtain the low frequency limit $G_0$.  However, this is problematic because the shear cycle simulation time becomes substantially larger than $t_{gel}$, such that it is possible for significant evolution of the gel structure to occur during the shear simulation.  This is especially true for $K=0.1$.  This aging effect will complicate the interpretation of the rheological response \cite{Zia2014}.

\section{DISCUSSION AND CONCLUSIONS}

This paper describes simulations of an irreversibly-bonding colloidal gel in which short-ranged forces hinder particle rotation. An angular potential $U_a$ is introduced, in which the rotation of bonded particles is controlled by sinusoidally varying terms. The number of minima in a complete rotation, $n$, and the barrier between minima $2K$ are parameters which control the extent to which bonded particles may rotate.  The effects of this potential on the structure, dynamics, and rheology of the simulated gels were determined as functions of these parameters over a range spanning from nearly barrierless rotation to very strongly hindered rotation.

Both $K$ and $n$ have a significant impact on the dynamics of aggregation and gelation and the morphology of the gels formed. At low $K$ particles are nearly free to rotate, and consequently gel properties are mostly independent of number of local minima $n$.  The clusters formed under such conditions are compact, and they aggregate to form coarse networks with thick strands and relatively high average coordination number.
At higher $K$ and $n$, significant barriers to particle rotation hinder the compactification and restructuring of clusters. This results in highly branched clusters which aggregate to form more diffuse space-spanning networks with lower coordination numbers, especially at higher $n$. 
The dynamics of gelation is governed by the competition between coarsening/compactification and cluster-cluster aggregation. Low $K$ and $n$ favor compactification and result in longer gelation times, while high $K$ and $n$ lead to lower-density clusters and shorter aggregation times.

In addition to visual inspection of gel structures, a variety of quantitative metrics were used to investigate gel structure, all of which support the basic picture just described. These include analyses of the structure factor, radial distribution, distribution of the number of bonds per particle, distributions of deviation from initial contact angle at bonding, and fractal dimensions. Gelation dynamics were likewise probed by measuring the mean number of bonds per particle and the number of clusters remaining as a functions of time.  At low $K$, all these metrics were largely independent of $n$.  A large degree of restructuring is indicated by a broad distribution of bond angles $\delta\theta$, and structure in the radial distribution at small length scales consistent with dense packing. The number of bonds formed per particle at the end of the gelation simulations was $3.5$ under the conditions studied, again independent of $n$.

At high $K$ barriers to particle rotation hinder cluster compactification and restructuring, increasingly so as $n$ is increased. In the bond angle distributions at high $K$ only the minimum corresponding to the initial contact angle is populated, indicating that particles do not escape from this potential well on the timescales simulated. The mean number of bonds per particle decreases with increasing $K$ and $n$, and is as low as $2.0$ for $K=2.0$ and $n=16$, similar to that observed in hard-sphere--type DLCA simulations.  At $K=2.0$ the characteristic length scale decreases with increasing $n$, though such a dependence is not observed at $K=0.75$.
The gelation time decreases systematically with increased hinderance to particle rotation; at $K=2.0$ and high $n$ gelation occurs approximately one order of magnitude faster than at $n=1$. This suggests that particle roughness is a significant variable for processing of commercial gel-based products.

The mass fractal dimension $d_f$ was found to be only weakly dependent on $K$ and $n$.  $d_f$ was found in the range $2.05-2.16$ for all systems except at $K=2.0$ and $n=16$, for which $d_f=1.99$. This quantity therefore appears to be insensitive to small changes in the potential $U_a$, though for large changes in the potential there are clearly effects. At $K=2.0$ $d_f$ decreases monotonically with increasing $n$, though a large change is only observed between $n=4$ and $n=16$.  This suggests that yet higher values of $n$ (and possibly $K$) are required in order to obtain fractal dimensions close to the $d_f = 1.78$ obtained in DLCA models that do not allow any restructuring.

Mechanical properties of the simulated gels were measured using non-equilibrium oscillatory shear simulations over a frequency range spanning the transition from elastic to viscous-dominated response.  Two trends were observed. First, for higher $K$ and $n$ the gel networks become stiffer and the frequency dependence of the elastic moduli $G'$ become weaker. Second, higher $K$ and $n$ increase the crossover frequency from elastic to viscous-dominated behavior.

$G''$ was found to scale according to either $\alpha^{0.5}$ or $\alpha^{0.53}$, with only small deviations observed at high and low frequencies.  At frequencies above crossover, $G'$ scaled according to either $\alpha^{0.5}$ or $\alpha^{0.35}$, depending on $K$ and $n$; the lower exponent was observed for higher $K$ and $n$, where rotation is strongly hindered.
At high $K$, the crossover frequency from solid-like to liquid-like behavior shifts to higher frequency and is strongly $n$-dependent.  The low-$K$ scaling exponent of $0.5$ is consistent with both previous simulation studies of systems without rotational barriers.  The dependence of the high-frequency scaling exponent on $U_a$ suggests that linear rheology measurements of this type can be a sensitive tool for characterizing non-central bonding interactions.  Our findings of the frequency-dependent $G'$ agree with recent experiments on how suspension rheology varies with surface characteristics \cite{Schroyen2019}.

Aging can affect the mechanical properties of particle gels both before and during mechanical tests.  The rheology data in this study was extracted for systems at the same `waiting time' $t_w$ (all gelation simulations were of the same duration.) We have tested with different $t_w$ and confirmed that the reported results are qualitatively insensitive to the waiting time. To reduce the effects of aging, new-bond formation was suppressed during shear simulations. Structural and rheological analysis of per-cycle data indicated that only in the system with $K = 0.75$ and $n = 16$ was there any appreciable effect of aging over the frequency range studied.  Finally, at $K=0.1$ and $K=0.75$ the frequency range studied did not extend low enough to obtain the low-frequency limit of the modulus; aging effects may also prevent this limit from being achieved in these systems even if much longer simulations at lower frequencies were attempted.

In real systems, DLCA-like behavior corresponding to the high-$K$-and-$n$ conditions simulated here is more often observed in colloids of very small particles (such as gold or silica nanoparticles {\cite{Weitz1984a,Manley2005a}}). This is consistent, because such particles have high surface roughness (relative to their diameter), while the nearly-barrierless conditions are typical of gels of large spherical particles such as polystyrene.  Mapping of real particle properties to the model parameters $K$ and $n$ is a nontrivial problem but could be approached either by empirical fitting to experimental rheological data or by using detailed atomistic simulations to study interparticle interactions and then coarse-graining.

  Several additional interesting questions remain open. In fractal particle gels many quantities scale with volume fraction. How those scalings change with $K$ and $n$ was not considered in the current study. The ``gel point'' (lowest $\phi$ at which connected structures form) was also not determined \cite{Manley2004}, which is of practical concern in applications where gels are used to stabilize consumer products, among others \cite{Burey2008}.
Future work will: i) examine gels formed via RLCA kinetics where more compact clusters expected to form (even in high $K$) would lead to different morphology, gel stability, and therefore having impact on the mechanical response. 
ii) investigate phenomena where bond breakage and reform are expected by a modification to include bond break. These studies are expected to elucidate the link between non-central bonding interactions and non-linear responses, e.g. large-amplitude shear \cite{Hsiao2012} or two--step yielding \cite{Chan2012}


%

\end{document}